\documentclass[aps,prb,twocolumn,showkeys,showpacs]{revtex4-1}
\usepackage{amssymb,amsmath,amsfonts,bm}
\usepackage{graphicx}
\usepackage{hyperref}
\usepackage{url}
\usepackage[dvipsnames]{xcolor}
\usepackage{float}
\begin{document}
\title{Critical behaviour in two-dimensional Coulomb Glass at zero temperature}

\author{Preeti Bhandari} 
\author{Vikas Malik}\email{vikasm76@gmail.com} 
\author{Syed Rashid Ahmad} 
\affiliation{Department of Physics, Jamia Millia Islamia, New Delhi 110025, India}
\affiliation{Department of Physics and Material Science, Jaypee Institute of Information Technology, Uttar Pradesh, India}

\date{\today}

\begin{abstract}
The lattice model of Coulomb Glass in two dimensions with box-type random field distribution is studied at zero temperature for system size upto $96^{2}$. To obtain the minimum energy state we annealed the system using Monte Carlo simulation followed by further minimization using cluster-flipping. The values of the critical exponents are determined using the standard finite size scaling. We found that the correlation length $\xi$ diverges with an exponent $\nu=1.0$ at the critical disorder $W_{c} = 0.2253$ and that $\chi_{dis} \approx \xi^{4-\bar{\eta}}$ with $\bar{\eta}=2$ for the disconnected susceptibility. The staggered magnetization behaves discontinuously around the transition and the critical exponent of magnetization $\beta=0$. The probability distribution of the staggered magnetization shows a three peak structure which is a characteristic feature for the phase coexistence at first-order phase transition. In addition to this, at the critical disorder we have also studied the properties of the domain for different system sizes. In contradiction with the Imry-Ma arguments, we found pinned and non-compact domains where most of the random field energy was contained in the domain wall. Our results are also inconsistent with Binder's roughening picture.
\end{abstract}


\maketitle

\section{Introduction}
\label{intro} Coulomb glass (CG) belongs to the class of disordered insulators. The electronic states in the system are localized due to disorder and are unable to screen the Coulomb interactions effectively at low temperatures. Some examples of such a system are amorphous semiconductors and impurity bands in doped semiconductors where the Fermi level lies in the region of localized states.

In the absence of disorder, M\"{o}bius and R\"{o}ssler \cite{Mob} have given evidence of phase transition from fluid to the charge-ordered phase (COP) in two and three dimensional CG. There finite-size scaling (FSS) analysis predicts that the values of the critical exponents are consistent with those of the Ising model with short-range interactions. But at finite disorder, such a transition was seen in three dimensional (3d) CG \cite{Mob,over,surer,pankov}. Later the phase diagram and the critical properties of 3d CG were also investigated by Goethe and Palassini \cite{Martin}. They found fluid to COP transition consistent with random field Ising model (RFIM) universality class. In our previous paper \cite{preeti} we provided numerical evidence of COP at finite disorder in two dimensional (2d) CG. The investigations were done at zero temperature, where we found that the transition was driven by rearrangement of domain wall in the metastable state of COP as disorder was increased to give disordered phase. On the basis of this a two valley picture was proposed and phase coexistence was argued. This coupled with jump in the staggered magnetization at transition region was found as an indication of first order transition.

In this paper we will investigate the critical properties of 2d CG using finite size scaling at zero temperature. We have used Monte Carlo (MC) annealing and cluster-flipping algorithm to obtain the minimum energy state. The aim of this paper is to provide sufficient evidence that the transition is, indeed first order. The distribution of staggered magnetization was investigated around the transition region to confirm the presence of coexisting phases. In addition to this, we have also investigated the properties of domains in the system at transition.

From the earlier studies done on RFIM one can extract some useful conclusions about the phase transition and the properties of domains. One of the pioneer work in this field was done by Imry and Ma \cite{Imry}. They proposed that, if the spins within a domain of linear size $L$ are reversed, then the energy cost $E_{c} \sim JL^{d-1}$ where $J$ is ferromagnetic interactions between the nearest neighbour spins and $d$ is the dimensionality of the system. Reversing the spins yields gain in energy by an amount $hL^{d/2}$ where h is the root-mean square deviation of the random fields. Hence the energy needed to form a domain of linear size L in d dimensions is 
\begin{equation}
\label{introeq1} E(L) \approx JL^{d-1} - hL^{d/2}
\end{equation}  
For $d \geq 2$, if $h \ll J$ then $E(L)$ is positive but for $d<2$, if $L$ is large enough then $E(L)$ becomes negative. So the ferromagnetic ordering is unstable when $d<2$. At $d=2$, both the terms in the r.h.s. of eq.~\ref{introeq1} are of the order $L$ so the argument is inconclusive. Later, Binder \cite{Binder} gave an argument that, if one allows for roughening of the domain walls then there exists a length scale $L_{b}$ given by 
\begin{equation}
L_{b} \propto exp[C (J/h)^{2}]
\end{equation}
where C is a constant $\mathcal{O}(1)$, for which $E(L>L_{b})$ becomes negative. So no long range order can exist for $d=2$. A rigorous proof was later provided by Aizenman and Wehr \cite{Aizen}, claiming absence of ferromagnetic ordering in 2d RFIM. Sepp\"{a}l\"{a} et al \cite{Sep,Seppal} numerically confirmed Binder's roughening arguments and thus absence of any phase transition in 2d RFIM. The existence of ferromagnetic ordering in 3d RFIM was confirmed at finite \cite{Bric,Imbrie} as well as at zero temperature \cite{Ak,Aa,Id,Og}. From the standard picture \cite{Aj,Ds,Tn,Jv} one finds that the transition is second order, but there are few arguments in support of first order transition as well \cite{Ah,Eb,Jm,Apy}. Recent work on 2d RFIM \cite{Frontera,Aiz,Spas,Suman} also claim presence of ordered state in the system at finite disorder, but the nature of transition is not clear. In our previous work \cite{preeti}, we have given an argument that for a compact domain, the energy of formation of domain in CG system also scales as $L^{d-1}$. We then used numerical data to show that the argument also holds for non-compact domains. We will here check whether the Imry-Ma arguments (on domain structure and random field energy of the domain) and the Binder's argument (on domain wall roughening) are valid for CG system or not. 

The rest of the paper is arranged as follows. In section~\ref{sec3} we will first discuss our model and then the numerical technique used to reach the minimum energy state. In section~\ref{sec5} results are shown where first a detailed study is done to prove discontinuity in magnetization and the critical exponent of correlation length ($\nu$) and magnetization ($\beta$) are computed. Then the calculation of disconnected susceptibility is used to prove that the condition for second-order transition is not satisfied. We have then shown the distribution of staggered magnetization where three peaks are present at the critical disorder for all system sizes. The $L_{b}$ scaling derived for RFIM \cite{Binder} is also discussed later for CG system. We have then studied the properties of the domains at transition where our results are in contradiction with the Imry-Ma arguments \cite{Imry}. Conclusions are presented in section~\ref{sec6}.
\section{Model and Numerical Simulation}
\label{sec3}
The lattice model of CG was first discussed by Efros and Shklovskii \cite{Efros}, where the states were assumed to be localized around centres, on a regular lattice of $L^{d}$ sites. For the case where the number of electrons are half the total number of sites in the lattice, the Hamiltonian can be considered as 
\begin{equation}
\label{H1} H = \sum_{i} n_{i} \phi_{i} + \frac{1}{2} \sum_{j \neq i} \frac{e^{2}}{\kappa |\overrightarrow{r_{i}} - \overrightarrow{r_{j}}|} (n_{i}-\frac{1}{2}) (n_{j}-\frac{1}{2})
\end{equation}
where $n_{i}$ denotes the electron occupation number which can take values 0 and 1, as the on-site Coulomb energy is assumed to be too large to permit more than one electron per site. The on-site energy $\phi_{i}$ at each site was considered as independent random variable with a probability distribution $P(\phi)$ defined as
\begin{equation}
\label{P1}  P(\phi)=\begin{cases}
    \frac{1}{2W}, & \text{if $\frac{-W}{2} \leqslant \phi \leqslant \frac{W}{2}$}.\\
    0, & \text{otherwise}.
  \end{cases}
\end{equation}
The width $W$ of this distribution characterizes the amount of disorderedness in the system. The distance ($r_{ij}=|\overrightarrow{r_{i}} - \overrightarrow{r_{j}}|$) between sites $i$ and $j$ was calculated using periodic boundary conditions (using minimum-image convention). The system under consideration possesses a particle-hole symmetry so the chemical potential $\mu=0$. All the energies are measured in units of $e^{2}/\kappa a$ where $a$ is lattice spacing $a$ and $\kappa$ is the dielectric constant of the medium. We are considering two-dimensional CG system on a square lattice. 

The minimum energy state was obtained using Monte carlo annealing. To start the simulation we used a completely random initial configuration $\lbrace s_{i} \rbrace$ (using Ising spin variable $s_{i}=n_{i}-1/2$) where half the sites were randomly assigned with $s_{i}=\frac{1}{2}$ and the remaining half with $s_{i}=\frac{-1}{2}$. In our previous work \cite{preeti},  $\lbrace \phi_{i} \rbrace$ were chosen in a correlated manner, i.e. for each run, $\lbrace \phi_{i} \rbrace$ were chosen from a box distribution $\lbrace -1,1 \rbrace$ and then multiplied by $W/2$ which was increased from 0 to 0.50 in small steps. In this paper at each $W$, $\lbrace \phi_{i} \rbrace$'s were chosen independently from a box distribution given by eq~\ref{P1}. Both type of simulations have been extensively done for RFIM. Metropolis algorithm \cite{Metro,Schr} was used which constitutes of random walk in space of all the possible configurations in the system. As the number of electrons is conserved, we have used Kawasaki Dynamics (spin-exchange) here. In this case a single Monte Carlo step (MCS) involves randomly choosing two sites $i$ and $j$ with opposite spins for spin exchange. If the state after spin exchange results into energy relaxation then the exchange is always done with the exchange probability
\begin{equation}
\label{P2} P_{ij} = 1
\end{equation} 
But if the above mentioned condition is not satisfied i.e. the spin exchange results into thermal excitation by an amount say $\Delta_{ij}$ then the exchange probability is
\begin{equation}
\label{P3} P_{ij} = exp \left\lbrace \frac{-\Delta_{ij}}{KT} \right\rbrace
\end{equation} 
where 
\begin{equation}
\label{P4} \Delta_{ij} = e_{j} - e_{i} -\frac{1}{r_{ij}}
\end{equation}
is the change in energy \cite{Efros} calculated using the single-particle Hartree energy ($e_{i}$)
\begin{equation}
\label{P5} e_{i} = \phi_{i} + \sum_{i \neq j} \frac{s_{j}}{r_{ij}}
\end{equation}
Annealing was done from $T=1$ to $T=0.01$ for different system sizes ($L=16,32,48,64,96$). At low temperatures the number of MCS were increased to a maximum of $5 \times 10^{5}$ at each temperature. For all system sizes, investigations were done from $W=0.0$ to $W=0.50$. 
\begin{figure}[t]
\includegraphics[width=0.99\columnwidth]{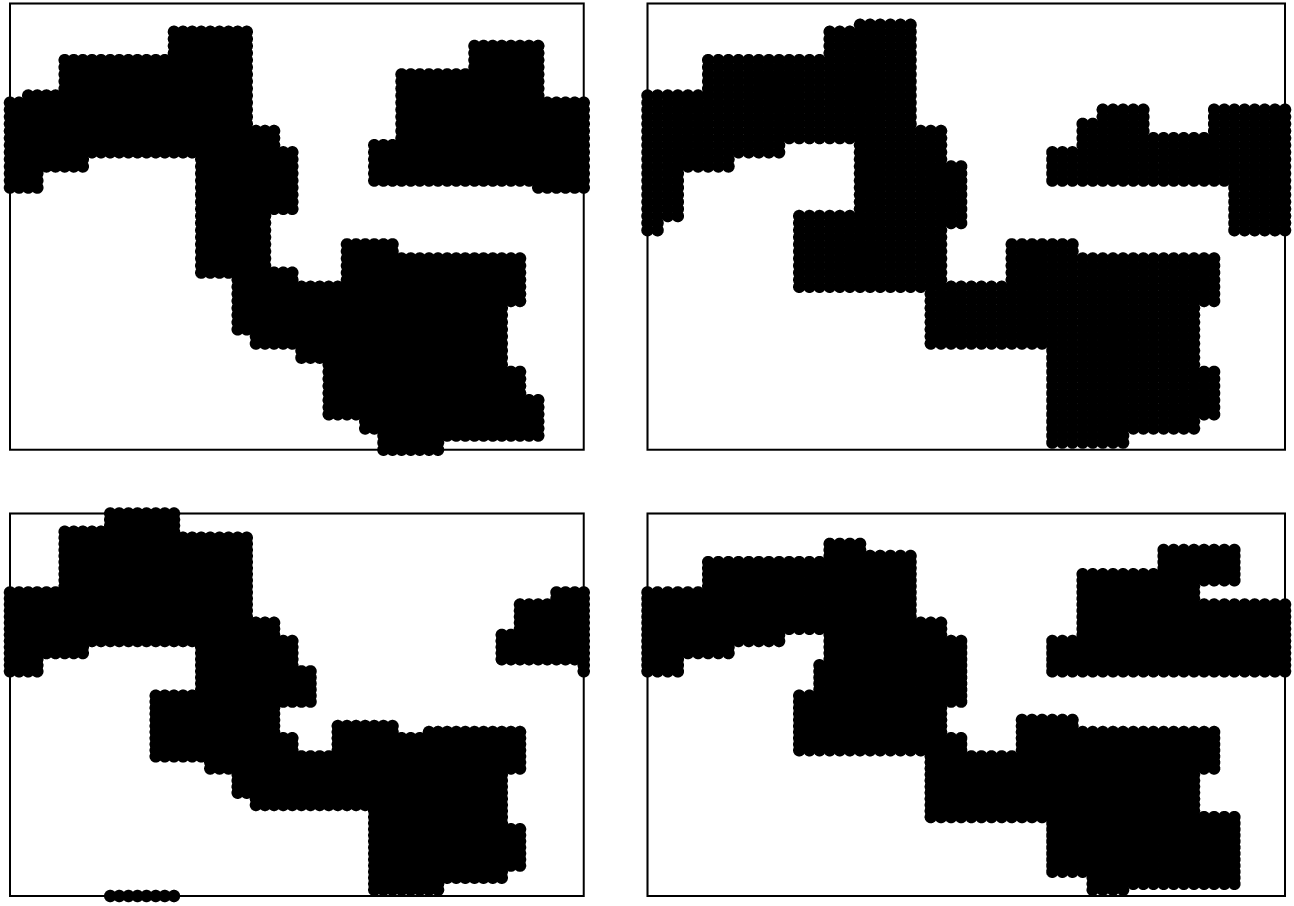}
\caption{Domains in the ground state of $L=64$ system having same disorder realization but four different initial configurations.} \label{fig1}
\end{figure}
The minimum energy state obtained after MC annealing was then used to perform cluster analysis by Hoshen-Kopelman algorithm \cite{Hoshen}. This algorithm is used to identify domains (clusters) which are defined as a group of nearest neighbour spins with antiferromagnetic ordering. We found that for $W \ll W_{c}$, the ground state consisted of a single domain. In the transition region, we found two large domains and few small domains. We then calculated the domain-domain interaction between all the domains excluding the largest domain. The interaction between these domains was negligible. So we flipped the domains one by one to reach a lower energy state. The final state thus obtained was then assumed as a ground state. We have also carried out the above mentioned simulation for the case where, same $\lbrace \phi_{i}\rbrace$ was considered for different initial configurations $\lbrace s_{i}\rbrace$ . The domains in the ground state for four different configurations are shown in Fig.~\ref{fig1}. One can see that the minimum energy states of all four configurations have the same domain structure and are pinned at a certain location. So using our method one cannot find the true ground state but the minimum energy state found will be very close to the ground state.
\begin{figure}
\includegraphics[width = 0.99\columnwidth]{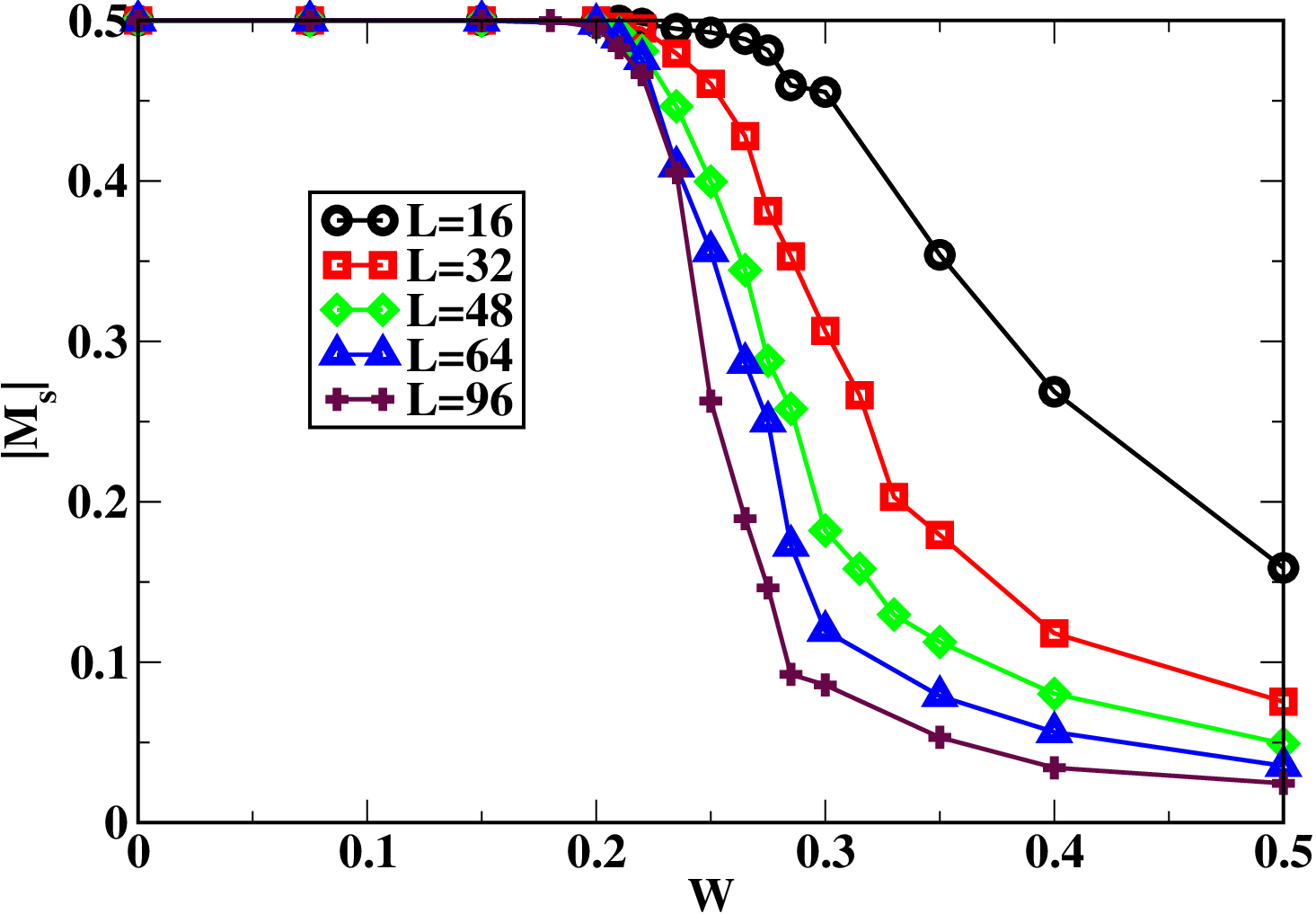}

\vspace*{9mm}
\includegraphics[width = 0.99\columnwidth]{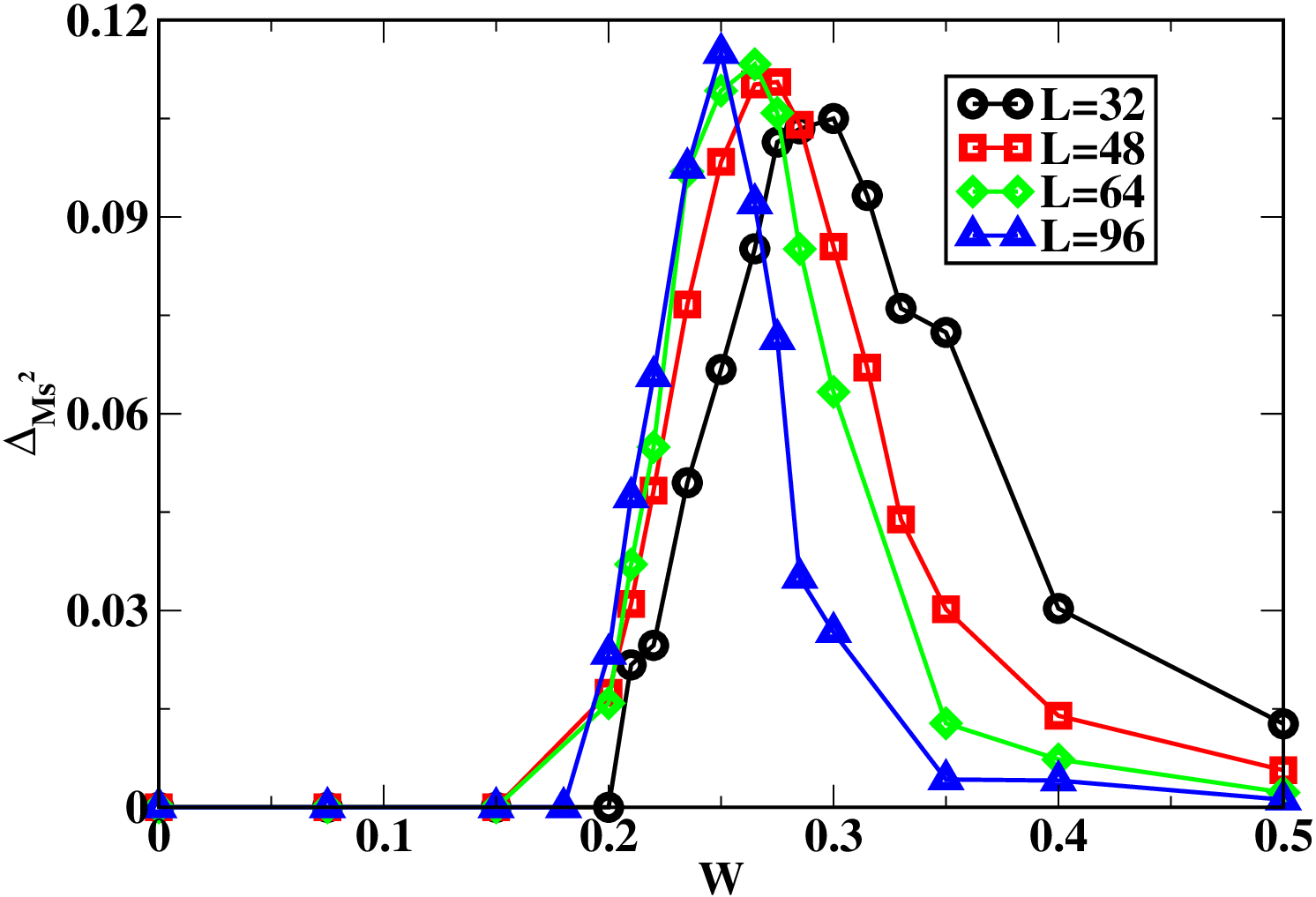}
\caption{(Color online) (Top) Behaviour of $|M_{s}|$ as a function of disorder. (bottom) Behaviour of $\Delta_{M_{s}^{2}}$ for different system sizes shows that the peak value of $L=96,64,48$ are saturating to a limiting finite value.} \label{msdata}
\end{figure}
\section{Results and Discussions}
\label{sec5}
\subsection{Calculation of Critical Exponents}
\label{sec5.1}

\begin{figure}
\includegraphics[width=0.99\columnwidth]{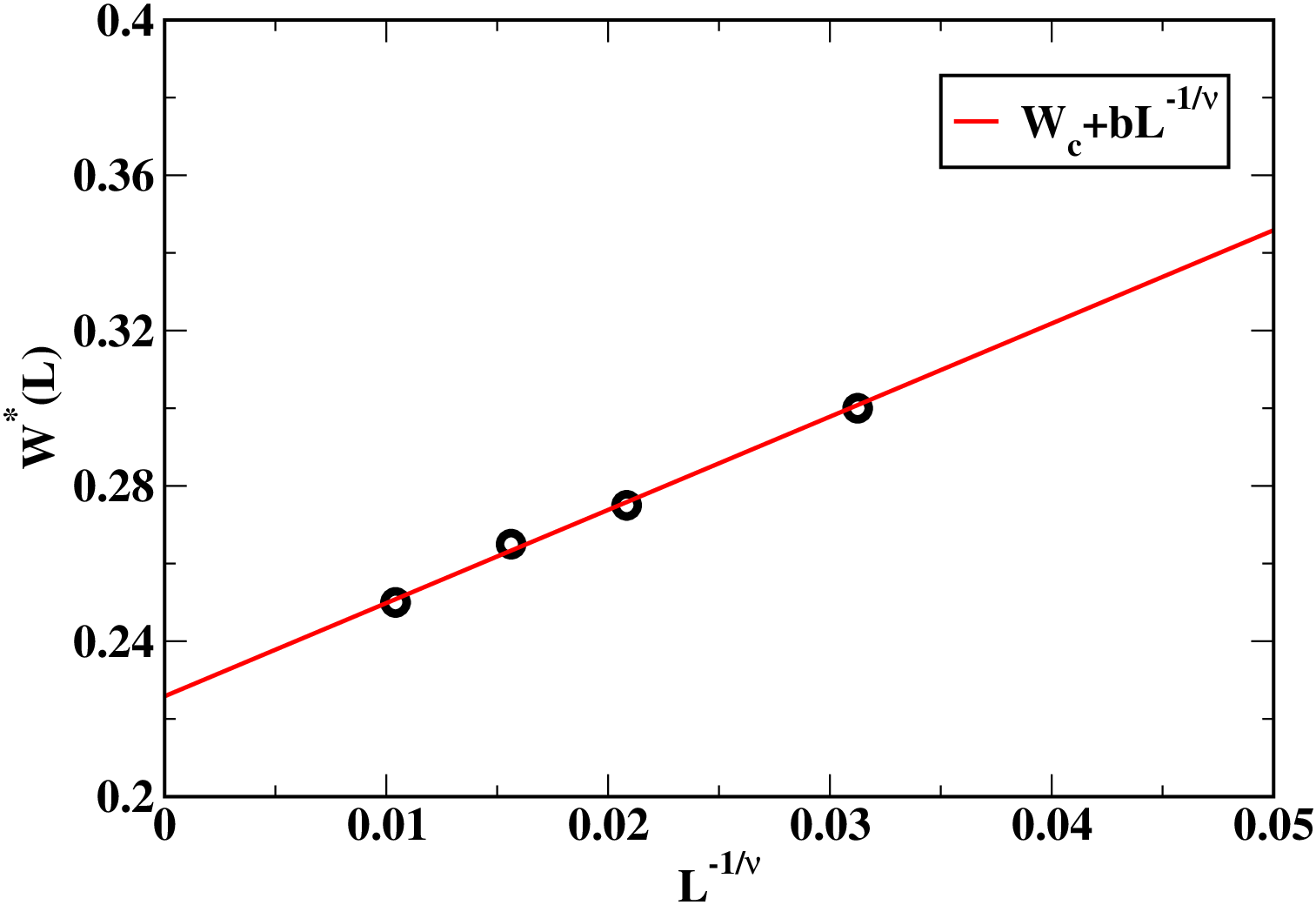}

\vspace*{1cm}

\includegraphics[width=0.99\columnwidth]{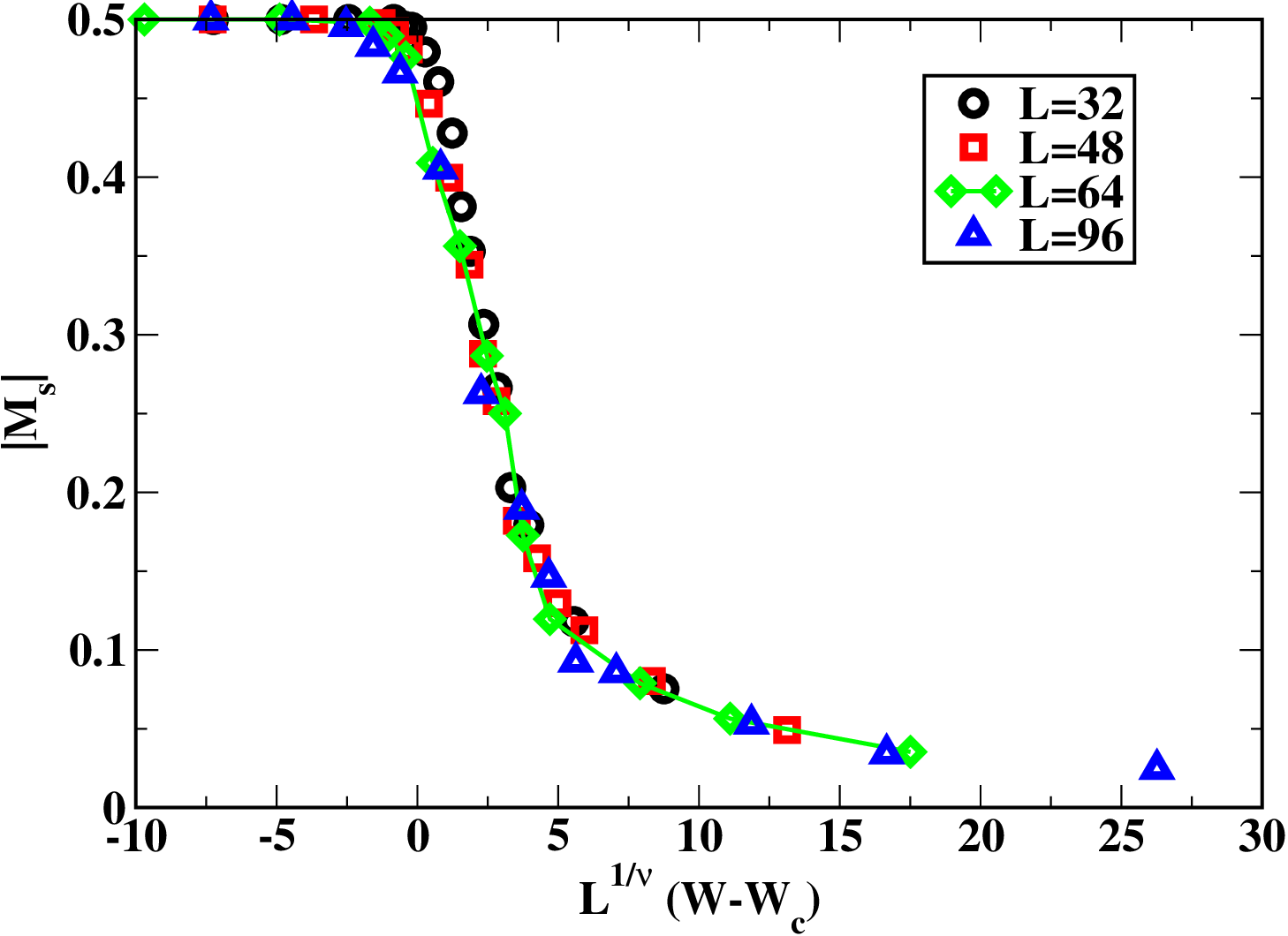}
\caption{(Color online) (Top) Plot of disorder $W^{*}(L)$ (where $\Delta_{M_{s}^{2}}$ attains its maximum) versus $L^{-1/\nu}$. Here we have not considered $L=16$ for scaling and the full line is a least-squares straight-line fit. (bottom) Scaled plot of the absolute value of average staggered magnetization using the parameters $W_{c}=0.22533$ and $1/\nu=1.0$ is shown. The full line connects the points for $L=64$ as a guide to the eyes. Here we have not rescaled $M_{s}$ by the factor $L^{-\beta/\nu}$, implying that $\beta=0$.} \label{msscaledata}
\end{figure}
In the absence of disorder, the ground state of CG is expected to have Anti ferromagnetic ordering \cite{Mob}. So the order parameter of the system is staggered magnetization, which is defined as follows
\begin{equation}
\label{eq1} M_{s} = \frac{1}{L^{d}} \sum^{L^{d}}_{i=1} \sigma_{i}
\end{equation}
where $\sigma_{i} = (-1)^{i} S_{i}$. Fig.~\ref{msdata}(top), shows the behaviour of $\langle |M_{s}| \rangle$ as a function of disorder, where $\langle...\rangle$ denotes disorder averaging. Here disorder averaging means averaging over different disorder $\lbrace \phi_{i} \rbrace$ realizations. In addition to this, we have also calculated the root-mean-square of different disorder realizations of the square of the staggered magnetization defined as follows \cite{Aa}. 
\begin{equation}
\label{eq2} \Delta_{M_{s}^{2}} \equiv \sqrt{\langle M_{s}^{4}\rangle - \langle M_{s}^{2}\rangle^{2}}
\end{equation}

One would expect that the peak of $\Delta_{M_{s}^{2}}$ to scale as $L^{-2\beta/\nu}$ but the peak values are slowly arriving to saturation as $L$ grows, which is shown in Fig.~\ref{msdata}(bottom). The data thus indicates that either $\beta=0$ or is a very small value which implies that magnetization is changing discontinuously. Similar results were obtained in 3d RFIM \cite{Aa}. To extract the location of transition, we determined the $W^{*}(L)$, where the maximum of the $\Delta_{M_{s}^{2}}$ occurs at each L and fit the correlation length exponent $\nu$ in such a way that our $W^{*}(L)$ values lie on a straight line if plotted against $L^{-1/\nu}$. The intercept of this straight line tells the value of the critical disorder ($W_{c}$). The fitted data points are shown in Fig.~\ref{msscaledata}(top) render 
\begin{equation}
\label{eq3} W_{c} = 0.22533 \hspace*{3mm} and \hspace*{3mm} \nu = 1.0.
\end{equation}
For a transition to be first order \cite{chayes}, the correlation length exponent $\nu$ should be equal to $2/d$, which is satisfied here. With the estimate for $W_{c}$ and $\nu$ from eq.~\ref{eq3}, the staggered magnetization data was then scaled using the standard finite-size scaling relation 

\begin{equation}
\label{eq4} \langle |M_{s}| \rangle = L^{-\beta/\nu} \hspace*{1mm} \tilde{M_{s}} ((W-W_{c})L^{1/\nu}) 
\end{equation}
As shown in Fig.~\ref{msscaledata}(bottom), we have not rescaled $\langle|M_{s}|\rangle$ by the factor $L^{-\beta/\nu}$, implying that $\beta=0$. This further confirms the discontinuity in staggered magnetization. Very small value of $\beta$ has been consistently found in all the studies of 3d RFIM at $T=0$ \cite{Aa,Id,Swift,Hart,New} as well as in $T\neq 0$ \cite{Reiger}, but this is not considered as a conclusive proof of a first order transition.

\begin{figure}
\includegraphics[width=0.99\columnwidth]{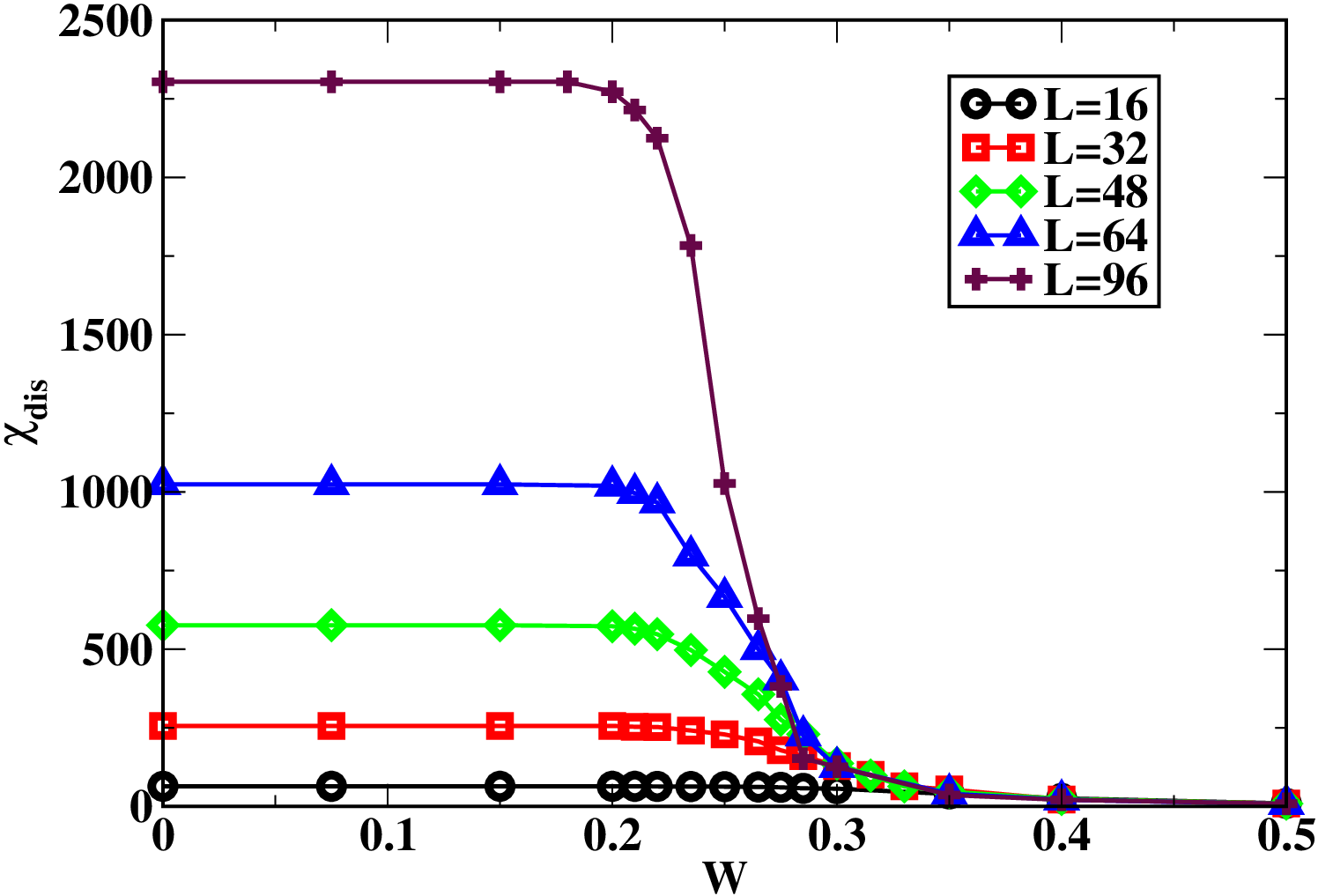}

\vspace*{1cm}
\includegraphics[width=0.99\columnwidth]{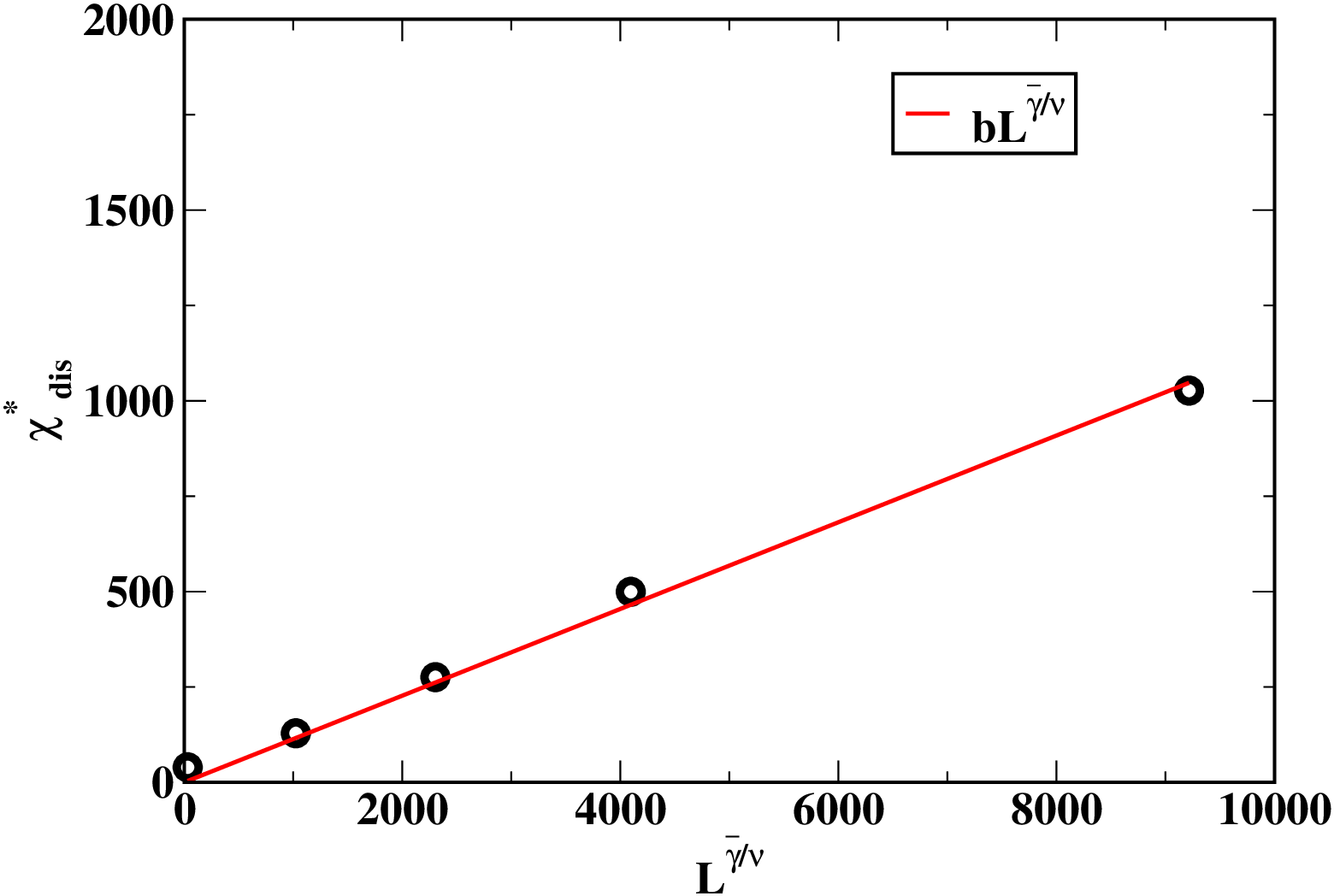}

\vspace*{1cm}
\includegraphics[width=0.99\columnwidth]{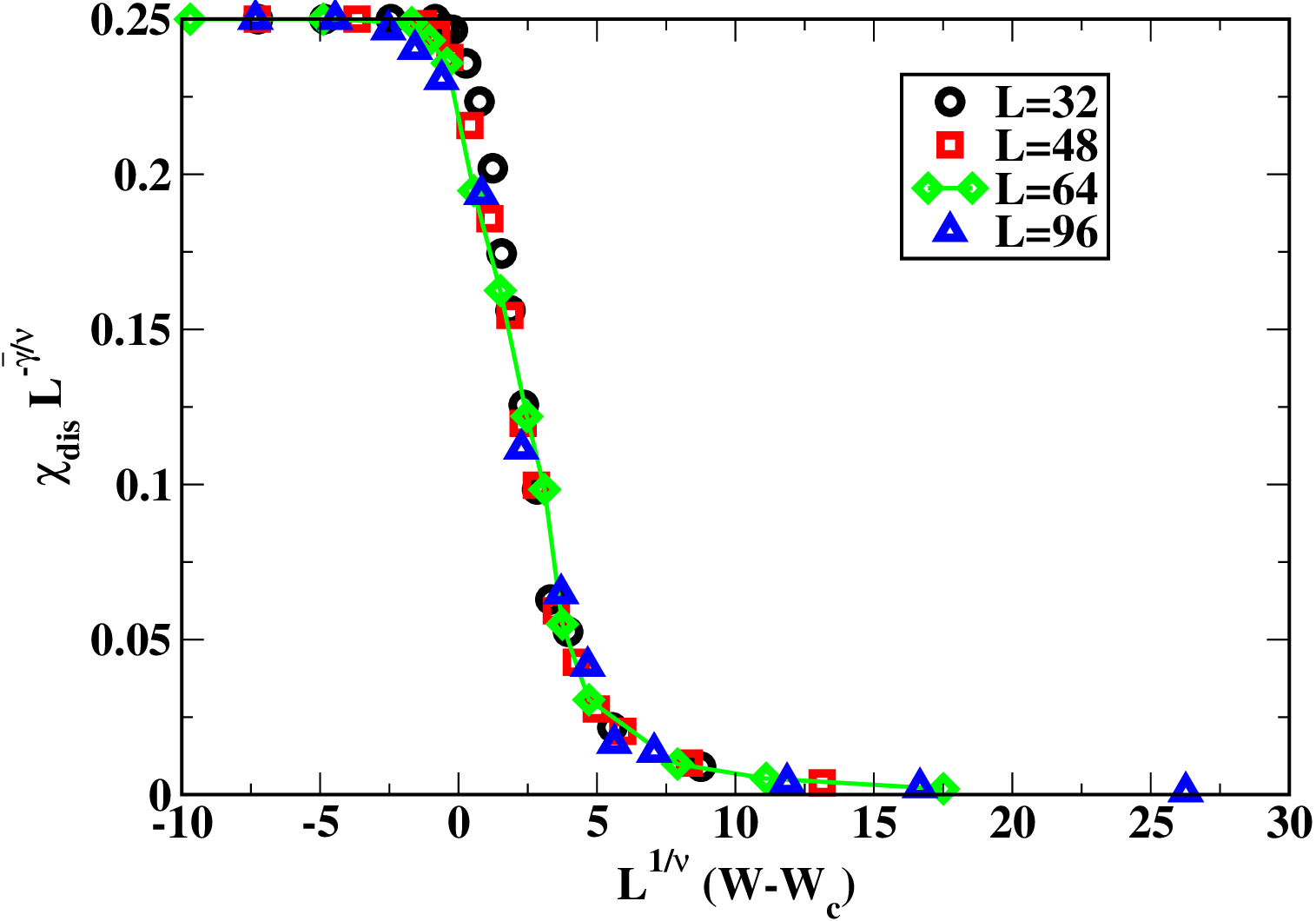}
\caption{(Color online) (top) Behaviour of $\chi_{dis}$ vs $W$. (center)The disconnected susceptibility value at $W^{*}(L)$ plotted as function of $L^{\overline{\gamma}/\nu}$,$\overline{\gamma}/\nu = 2.0$. (bottom) One can see scaling plot of the disconnected susceptibility with the parameters $W_{c}=0.22533$,$1/\nu=1.0$ and $\overline{\gamma}/\nu = 2.0$. The full line connects the points for $L=64$ as a guide to the eyes.} \label{Xdisdata}
\end{figure}

From a scaling argument \cite{Apy}, one can find that, a transition cannot be second order unless the following relation is satisfied
\begin{equation}
\label{eq5} d - 4 + \overline{\eta} > 0
\end{equation} 
To check whether this condition is satisfied here or not, we calculated disconnected susceptibility defined as
\begin{equation}
\label{eq6} \langle \chi_{dis} \rangle = L^{d} \langle M_{s}^{2}\rangle
\end{equation}
Fig.~\ref{Xdisdata}(top) shows the behaviour of $\langle \chi_{dis} \rangle$ at different disorders for all L. The value of $\langle \chi_{dis} \rangle$ at $W^{*}(L)$  denoted by 
$\chi^{*}_{dis}$  scales as 
\begin{equation}
\chi^{*}_{dis} \sim L^{\overline{\gamma}/\nu}
\end{equation}
$\chi^{*}_{dis}$ is plotted against $L^{\overline{\gamma}/\nu}$ and as shown in Fig.~\ref{Xdisdata}(center), the points are scaled along a straight line giving
\begin{equation}
\label{eq7} \overline{\gamma}/\nu = 2.0
\end{equation}
These exponents are related to $\overline{\eta}$ as \cite{Reiger} 
\begin{equation}
\label{eq8} \frac{\overline{\gamma}}{\nu} = (4-\overline{\eta})
\end{equation} 
From eq.~\ref{eq7} and eq.~\ref{eq8}, we get $\overline{\eta}=2$. Hence eq.~\ref{eq5} is not satisfied, which implies that the transition is not a second-order transition. Similar conclusion was drawn from 3d RFIM results \cite{Apy}. From the values of critical exponents at eq.~\ref{eq3} and ~\ref{eq7}, $\langle \chi_{dis} \rangle$ was scaled using the standard finite-size scaling relation (shown in Fig.~\ref{Xdisdata}(bottom).)
\begin{equation}
\label{eq9} \chi_{dis} = L^{\overline{\gamma}/\nu}  \hspace*{1mm} \tilde{\chi}_{dis} ((W-W_{c})L^{1/\nu})
\end{equation}
We here summarize that both disconnected susceptibility per site and staggered magnetization are independent of system size at critical disorder. 
%
\subsection{Distribution of staggered magnetization}
\label{sec5.2}
\begin{figure}
\includegraphics[width=0.99\columnwidth]{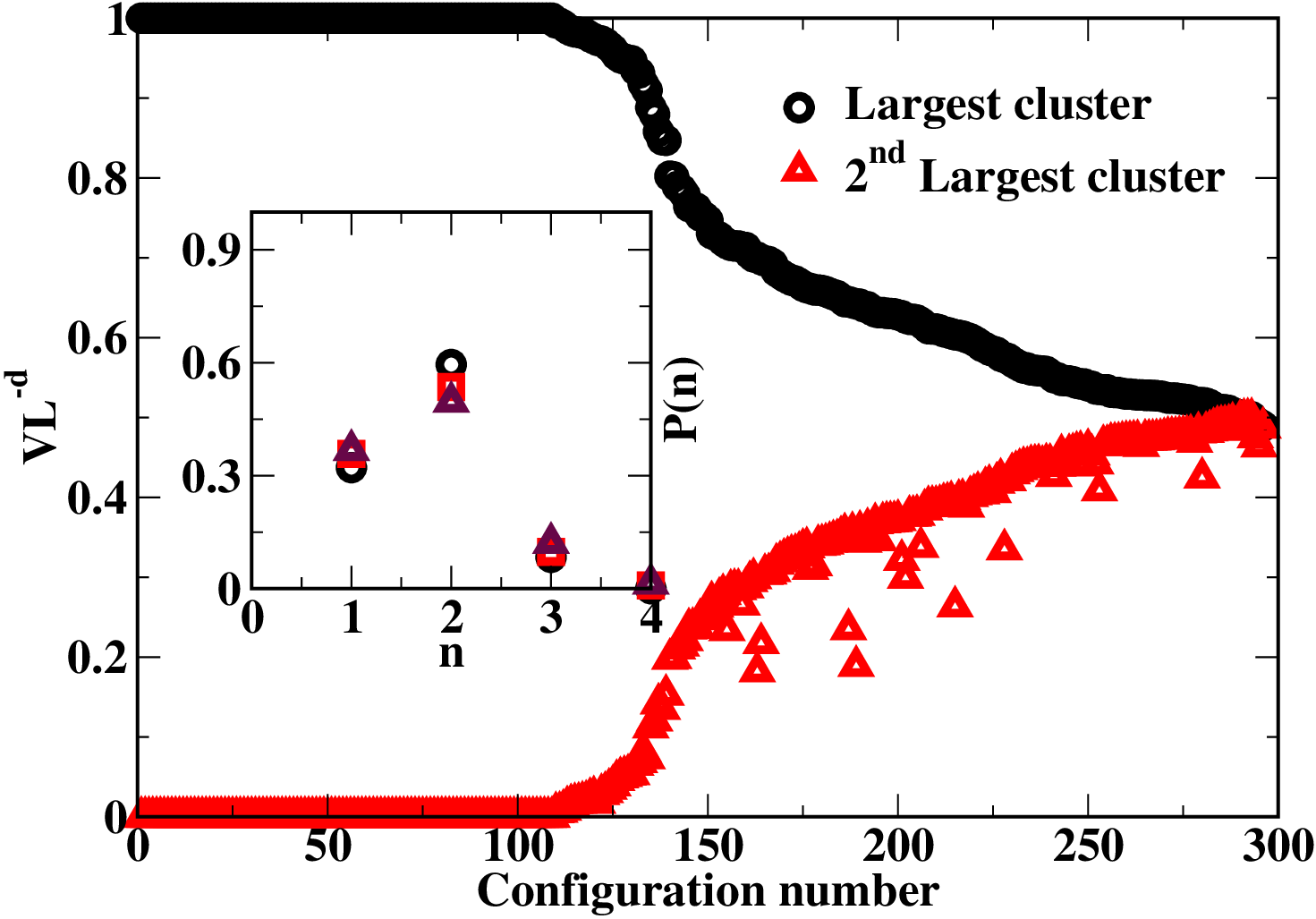}
\caption{(Color online) Largest and second largest domain size ($V$) at ($W_{c}=0.265$) divided by the system size $L^{d}=64^{2}$. The largest domains are sorted in descending order. Inset shows the probability ($P(n)$) of average number of domains(n) at respective $W_{c}$'s for $L=32 (\circ)$,$48$ ($\square$) and $64$ ($\triangle$)} \label{nodomain}
\end{figure}

\begin{figure}[b]
\vspace*{3mm}
\includegraphics[width=0.99\columnwidth]{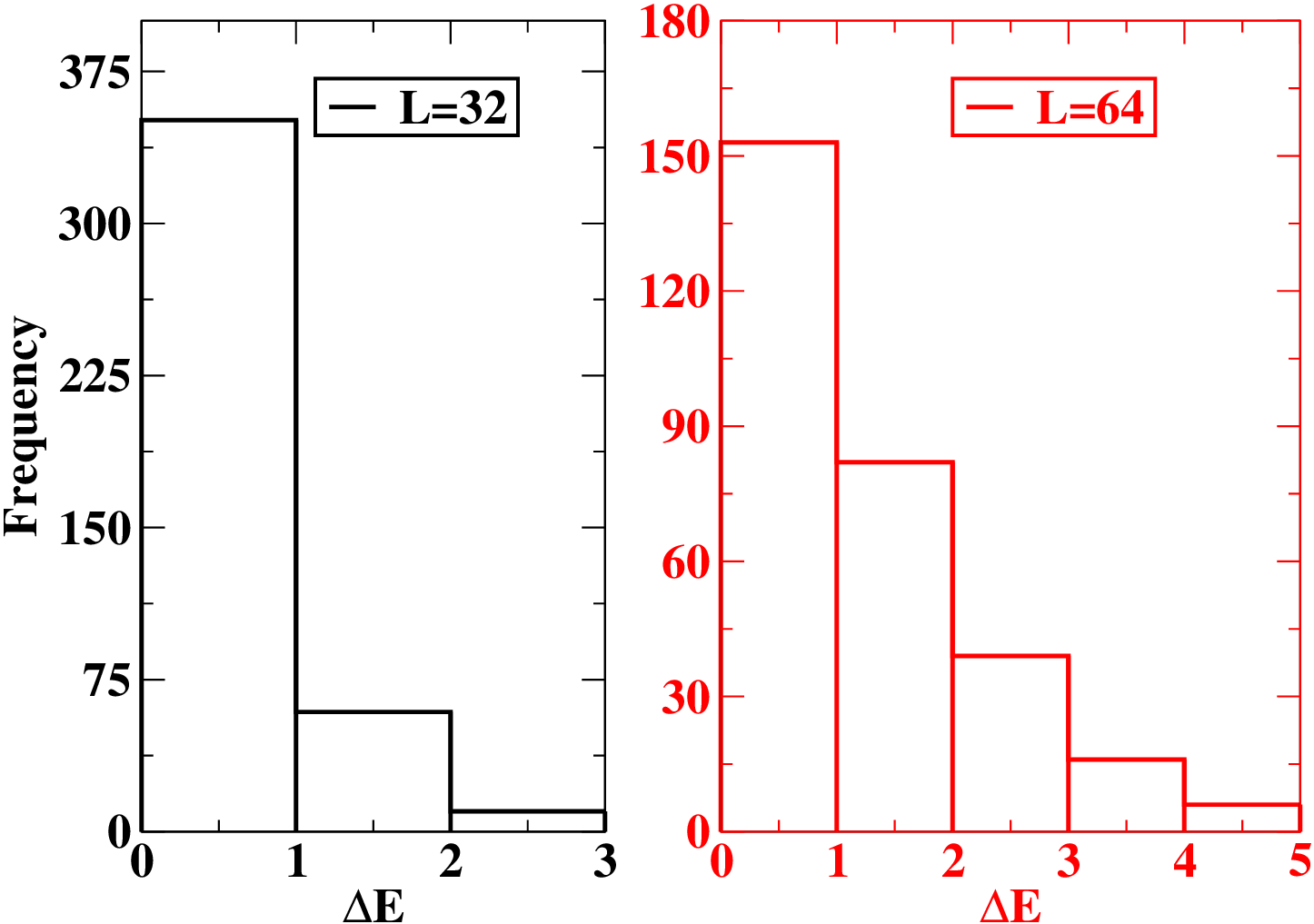}
\caption{(Color online) The energy difference ($\Delta E$) between the ground state and the domain state for $L=32$ and $L=64$ at there respective $W_{c}$'s. With increase in $L$, change in energy per site ($\Delta E/L^{d}$) is negligible.} \label{ediff}
\end{figure}

At $W_{c}$, ground state of the system was either COP or consisted of two large domains. As mentioned in section~\ref{sec3}, the minimum energy state after annealing consisted of several small domains and two large domains.  Cluster flipping led to small domains being flipped and energy of the system decreasing. The final state was mostly two large domains (domain state), which either got flipped leading to COP else to a disordered phase. The distribution of number of domains at transition and the size distribution of two largest domains is shown in Fig.~\ref{nodomain}. The energy difference between COP and domain state was very small as shown in Fig.~\ref{ediff}. When COP is the ground state, the domain state is the metastable state and vice versa.

\begin{figure}
\includegraphics[width=0.99\columnwidth]{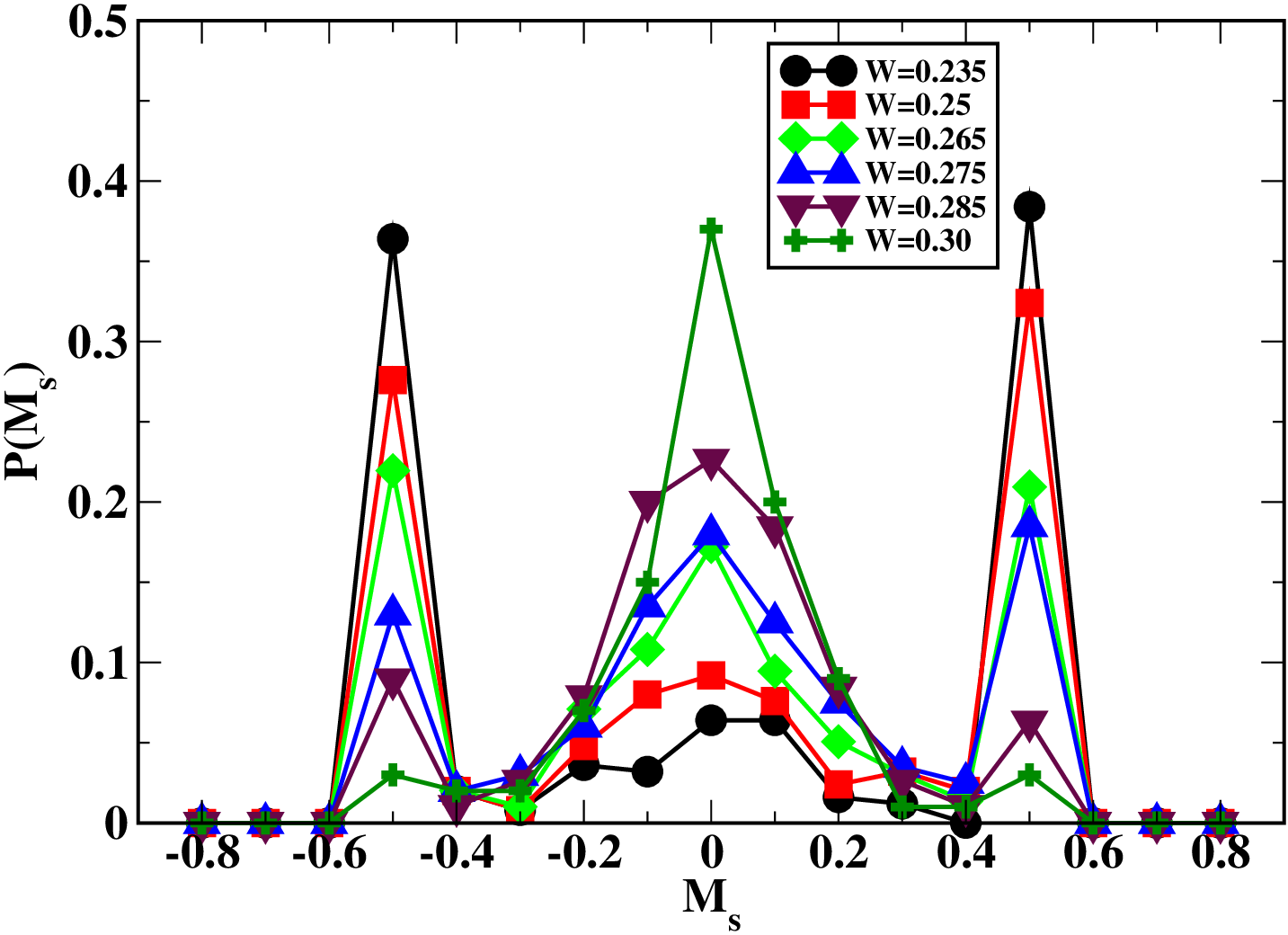}

\vspace*{1cm}

\includegraphics[width=0.99\columnwidth]{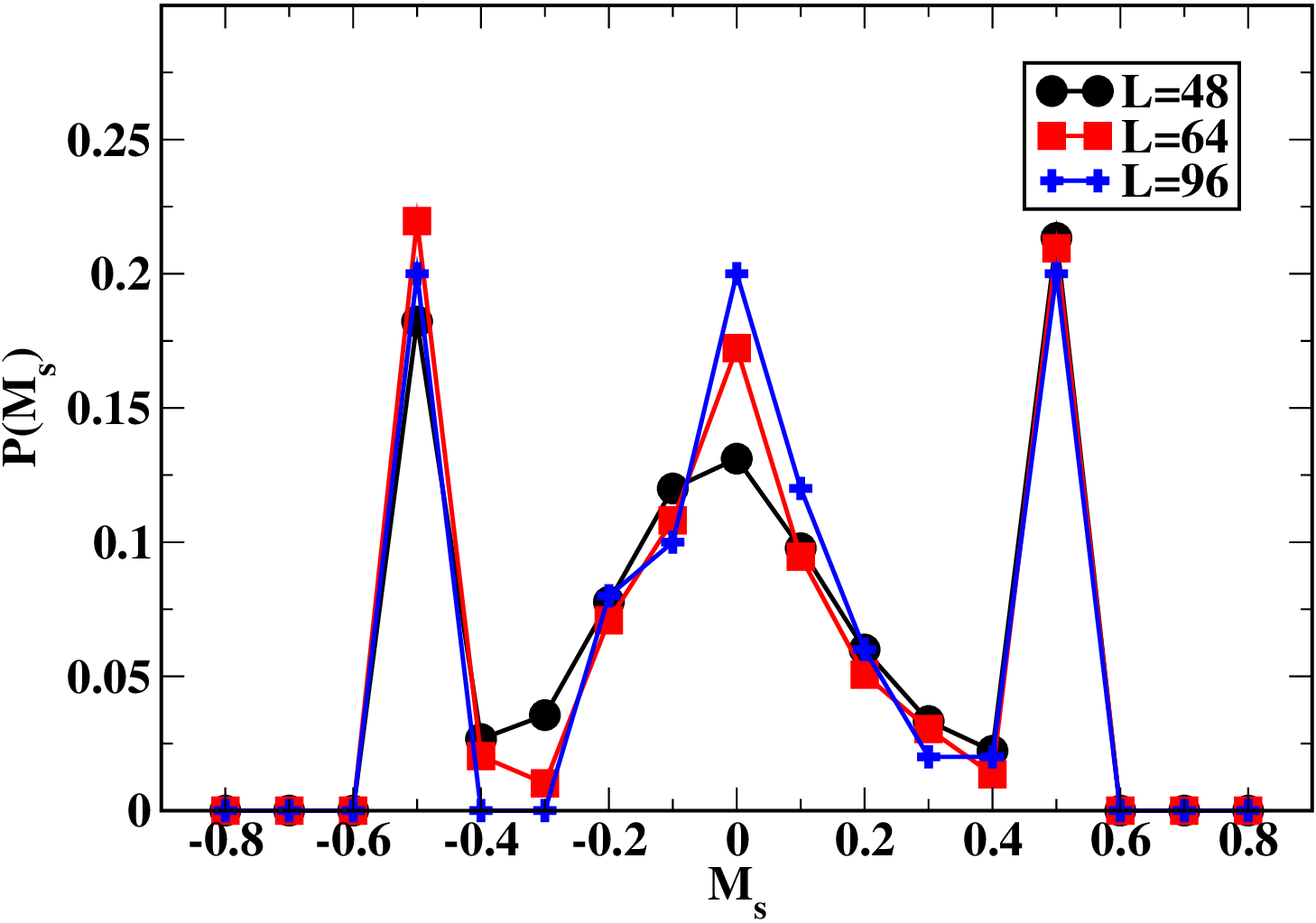}
\caption{(Color online) (top) Distribution of average staggered magnetization around the transition region for $L=64$. The distribution shows a three peak structure at $W=0.265$, which is $W_{c}$ for $L=64$. (bottom) Distribution of average staggered magnetization at respective $W_{c}$ of $L=48,64,96$.} \label{msLdistdata}
\end{figure}
In our previous work \cite{preeti}, we proved phase coexistence in terms of similar domains formed in the ground state at $W_{c}^{+}$ and in the metastable state at $W_{c}^{-}$. This implies that the free energy which is equal to energy at $T=0$ has two minimas (valleys) centred at $|\sigma|=0.5$ and $|\sigma|\approx small$, which is an indication of first order transition. To look for evidence of first-order transition, it is useful to plot staggered magnetization distribution $(P(M_{s}))$ vs $M_{s}$ in the transition region. One expects a three peak structure at $W_{c}$ signifying phase coexistence. In that case, the central peak corresponds to the disordered phase (where $\sigma \approx 0.0$) and the two side peaks corresponds to the charge-ordered phase (where $\sigma = \pm 0.5$). Also with increase in system sizes, the three peaks would become higher and sharper. Fig.~\ref{msLdistdata}(top) shows the distribution $(P(M_{s}))$ vs $M_{s}$ at various disorder for $L=64$. One can notice three peaks present in the vicinity of the critical disorder (which is $W_{c}=0.265$ for this system size). From Fig.~\ref{nodomain}, one can see that small domains are also possible. These are responsible for the finite width of $M_{s}=\pm 0.5$ peaks. Similarly for large domains, there is a wide distribution of domain sizes leading to a broad peak around $M_{s}=0$. For $W<W_{c}$ the distribution possesses two side peaks and as $W$ becomes greater than $W_{c}$ a single peak centred at $M_{s}=0$ becomes dominant. In Fig.~\ref{msLdistdata}(bottom), we have shown the distribution for $L=48,64,96$ only at there respective $W_{c}$, to confirm the presence of three peaks at all $L$. Getting three peaks structure at $W=W_{c}$ is not surprising. The energy difference between COP and the domain state is very small. At $W=W_{c}^{-}$, for most of the configurations, ground state is COP and the domain state is the metastable state. At $W=W_{c}^{+}$, the situation is reversed and at $W=W_{c}$, phase coexistence occurs. To summarize, the distribution of staggered magnetization suggests that the 2D CG undergoes a first-order transition as a function of disorder at $T=0$.

\subsection{Properties of the domain}
\label{sec5.3}

\begin{figure}
\includegraphics[width=0.99\columnwidth]{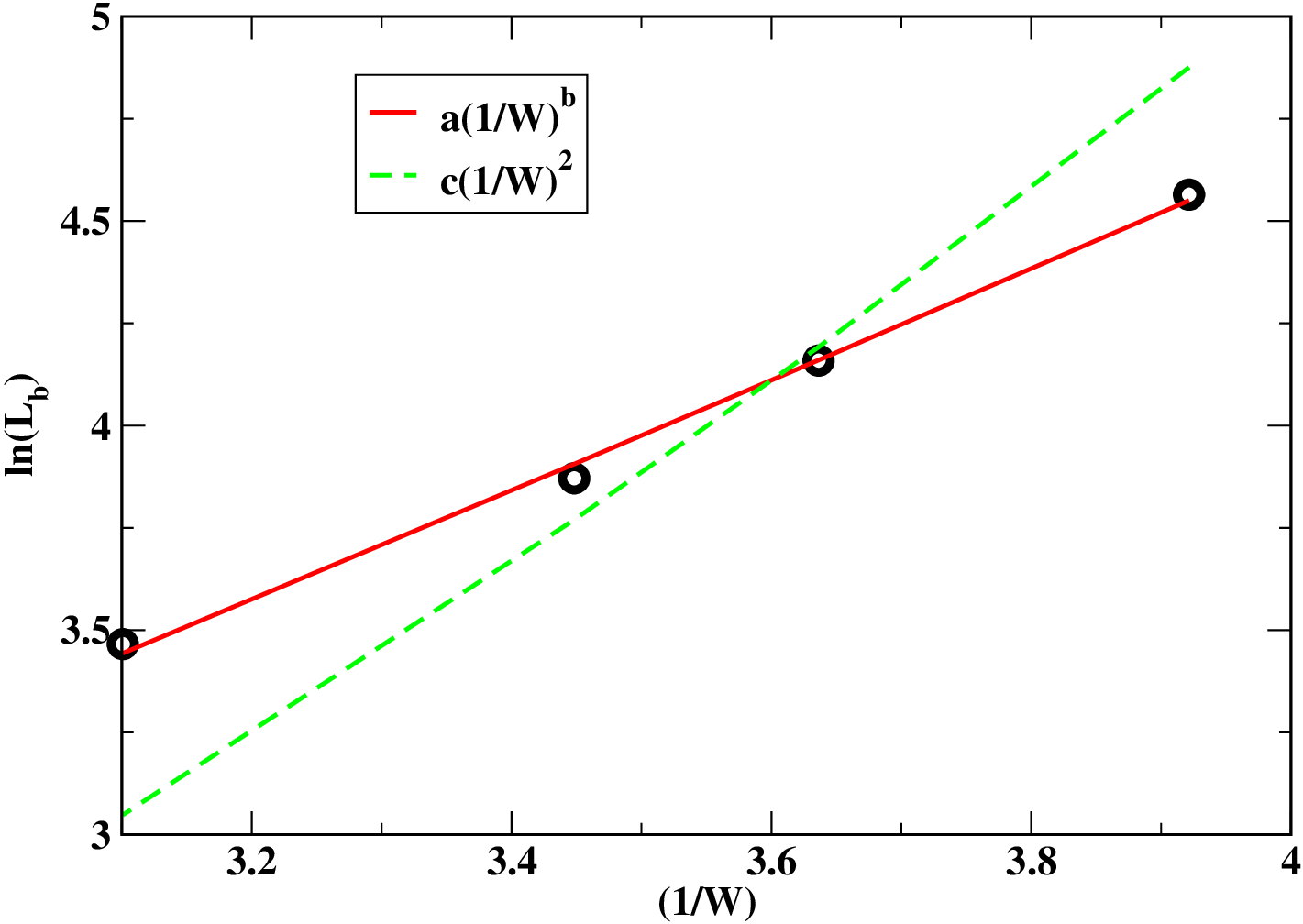}
\caption{(Color online) ln($L_{b}$) vs $(1/W)$, where $L_{b}$ is calculated using $P_{M_{s}}(L_{b}) = 0.25$. The solid line is the fitting using the function $f(x) = (a[1/W]^{b})$, where $a=0.89$, $b=1.18 $ and the dotted line is drawn using the function $f(x) = (c[1/W]^{2})$ where $c=0.31 \pm 0.01$ } \label{Lbwork}
\end{figure}

The absence of ferromagnetic ordering in 2D RFIM is due to the roughening of the domain walls. Hence there exists a breakup length $L_{b}$ above which the magnetization vanishes. This breakup length is related to the disorder in the system by the relation \cite{Binder}
\begin{equation}
\label{Lb1} L_{b} \sim exp(C[1/W]^{2})
\end{equation}
where C is a disorder-dependent constant of $\mathcal{O}(1)$. The variation of $L_{b}$ for varying W is shown in Fig.~\ref{Lbwork} where $L_{b}$ was defined with $P_{M_{s}}(L_{b})=0.25$. The data was scaled using the relation 
\begin{equation}
L_{b} \sim exp(a[1/W]^{b})
\end{equation}
Compared to the Binder's relation where $b=2$ (fitting shown with dotted line in Fig.~\ref{Lbwork}) our data fits well with $a=0.89$ and $b=1.18$. Hence the Binder's relation in eq.~\ref{Lb1} is not satisfied here although the possibility of exponential divergence in $L_{b}$ cannot be excluded.

We have then investigated the Imry-Ma arguments (discussed in section~\ref{intro}) on CG model. From the numerical studies on RFIM \cite{Cambier,Esser} one finds that the domains are pinned and non-compact and the random field energy of the domains exceeds considerably from the value calculated from the rms random field fluctuations. We first tested the compactness of the domain in our system at $W_{c}$ using the power law relation \cite{Cambier}
\begin{equation}
\label{S1} S \approx V^{\tau}
\end{equation} 
where $S$ denotes the total number of sites on the domain wall, $V$ is the total number of sites in the domain and $\tau$ is the surface exponent which for a compact domain takes the value 
\begin{equation}
\tau = \frac{d-1}{d}
\end{equation}
\begin{figure}
\includegraphics[width=0.99\columnwidth]{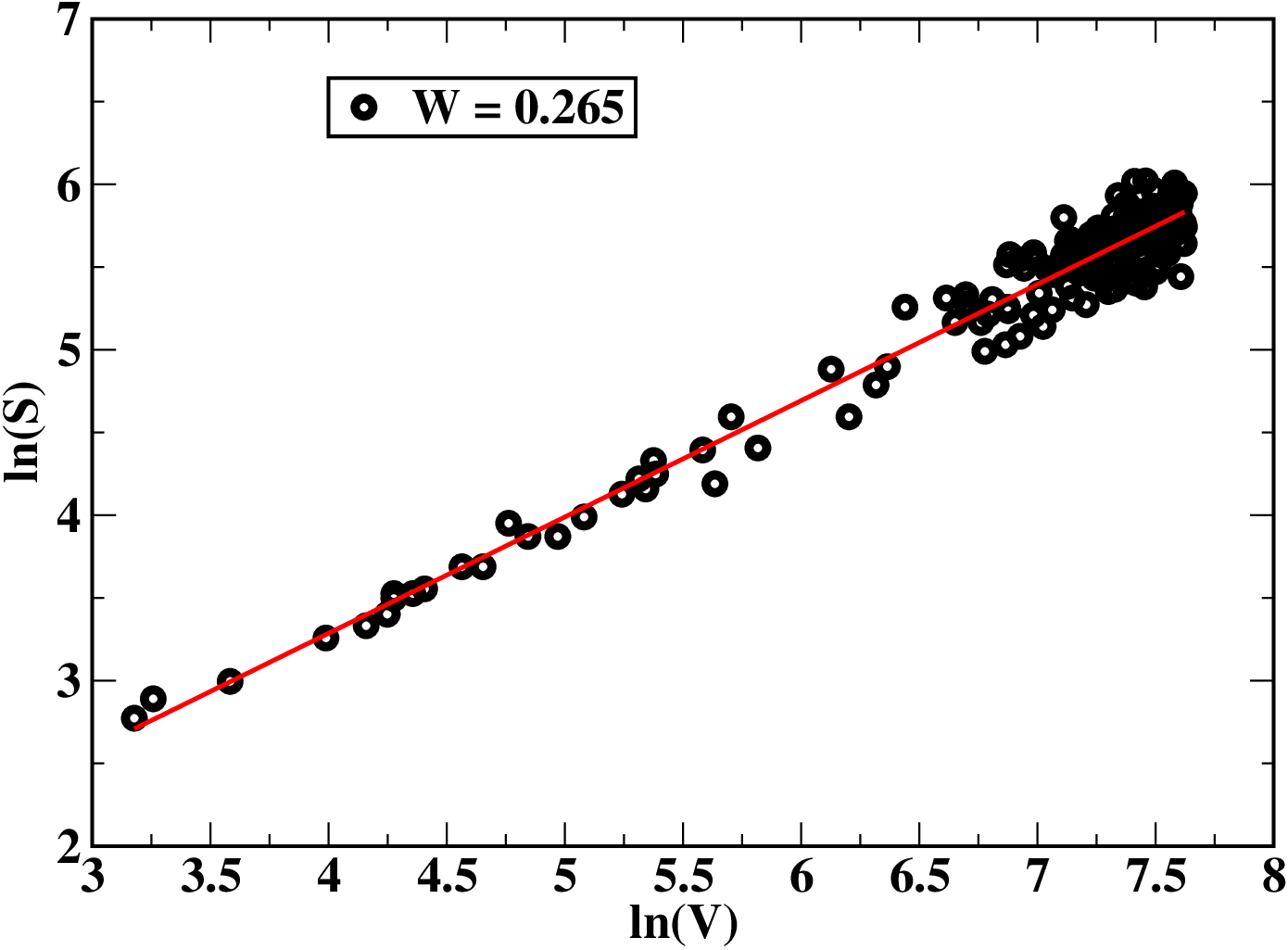}
\caption{(Color online) Logarithmic plot of $S$ vs $V$ for $L=64$ at its critical disorder $W_{c}=0.265$. From the power law relation in eq.~\ref{S1} we found $\tau=0.7030$. } \label{SVwork}
\end{figure}
\begin{figure}[b]
\vspace*{3mm}
\includegraphics[width=0.99\columnwidth]{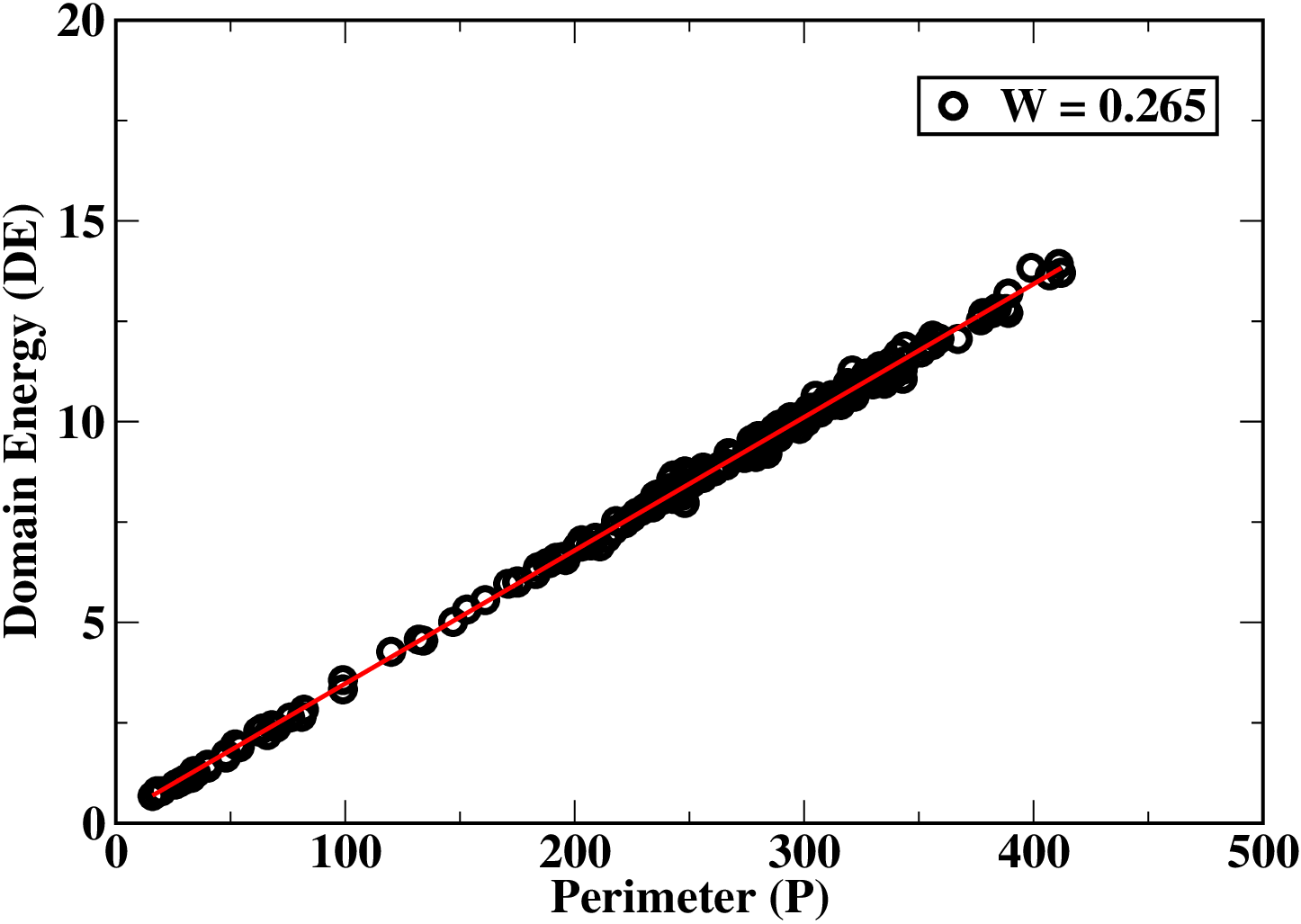}
\caption{(Color online) Domain energy ($DE$) vs perimeter ($P$) for $L=64$ at its critical disorder $W_{c}=0.265$ is plotted.} \label{DPwork}
\end{figure}

The scaled plot for $L=64$ is shown in Fig.~\ref{SVwork}. The value of $\tau$ is greater than $1/2$, for all system sizes as shown in Table~\ref{T1}, which indicates that the domains are non-compact.

For a short range system, the domain energy is proportional to $L^{d-1}$ as seen in first term of eq.~\ref{introeq1}. In our previous paper \cite{preeti}, we found that the domain energy (DE) was related to the perimeter (P) of the domain by the relation
\begin{equation}
DE \propto P
\end{equation}
The validity of the relation for $L=64$ is shown in Fig.\ref{DPwork}. So the first term in eq.~\ref{introeq1} defining the Imry-Ma argument holds.

\begin{figure}
\includegraphics[width=0.99\columnwidth]{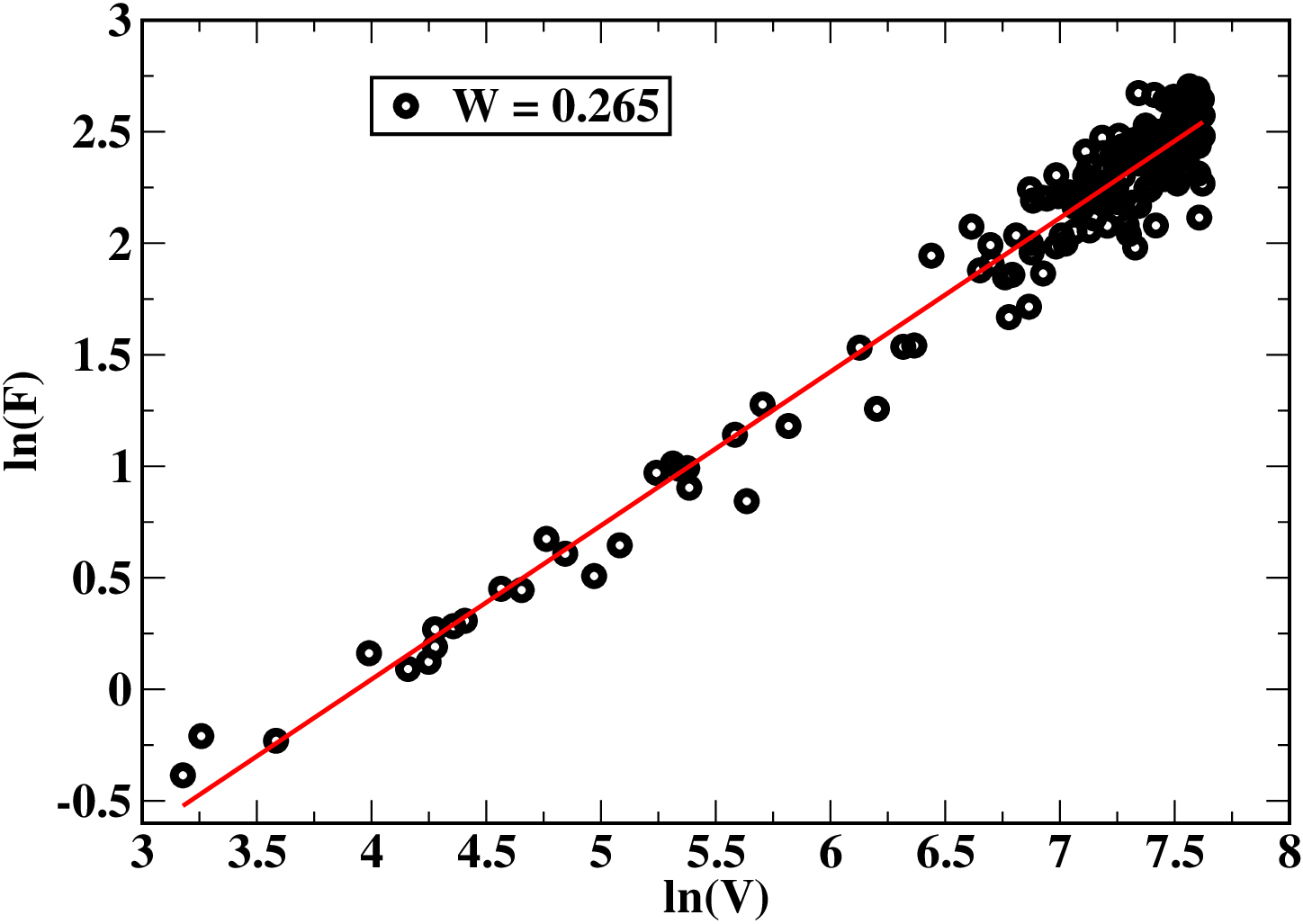}
\caption{(Color online) Logarithmic plot of $F$ vs $V$ for $L=64$ at its critical disorder $W_{c}=0.265$. From the power law relation in eq.~\ref{S2} we found $\lambda=0.6896$. } \label{FVwork}
\end{figure}

Whether the total random-field fluctuation ($F$) in a domain is proportional to the square root of $V$ or not was then tested using the power law relation \cite{Cambier}
\begin{equation}
\label{S2} F \approx V^{\lambda}
\end{equation}
where $\lambda$ was considered as an undetermined exponent. The slope of the data (shown in Fig.~\ref{FVwork}) predicts $\lambda$ to be significantly higher than $1/2$ (which was the theoretical value predicted by Imry and Ma). 
\begin{figure}[b]
\vspace*{3mm}
\includegraphics[width=0.99\columnwidth]{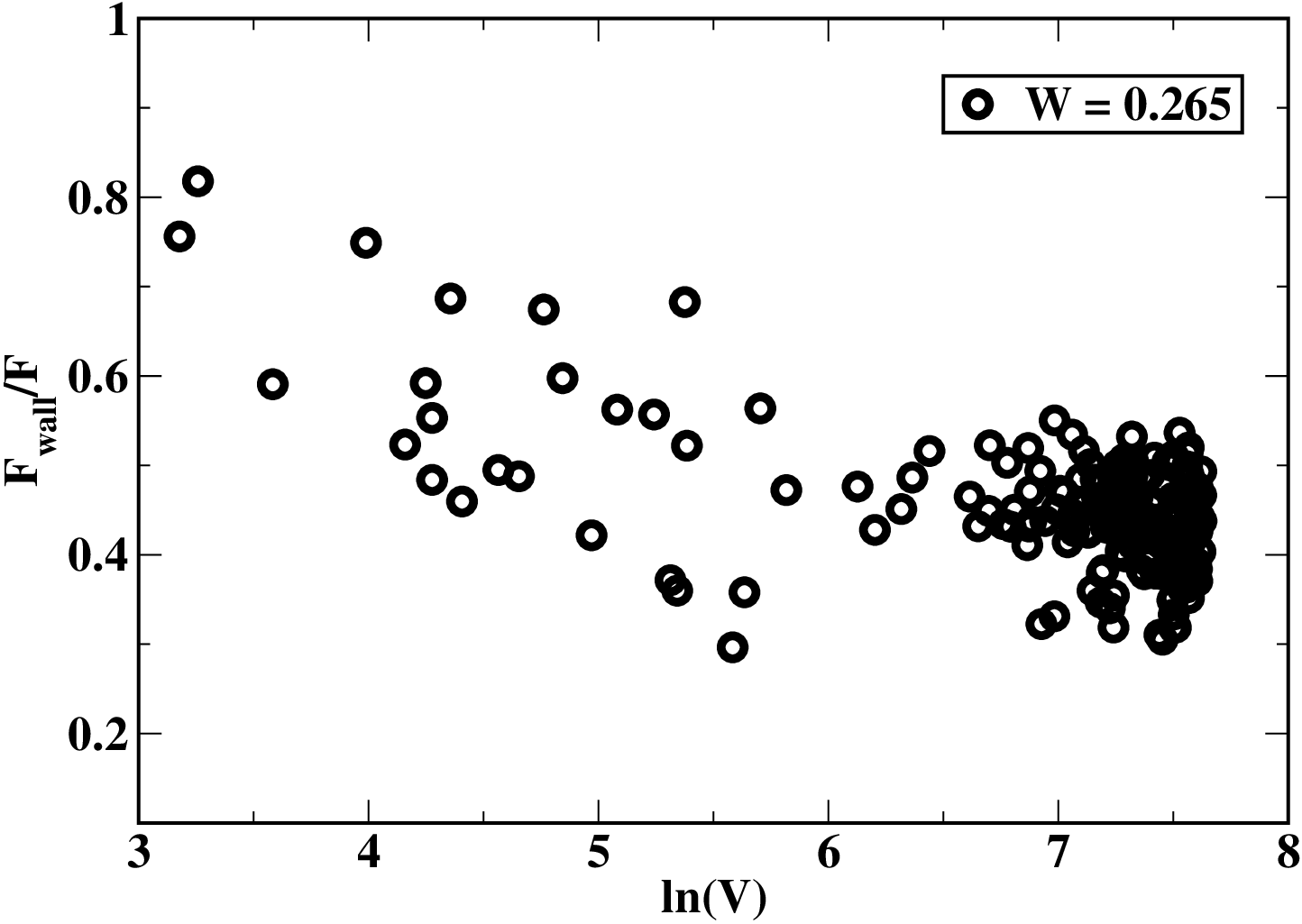}
\caption{(Color online) $F_{wall}/F$ vs $V$ for $L=64$ at its critical disorder $W_{c}=0.265$.} \label{ratiowork}
\end{figure}

\begin{figure}
\includegraphics[width=0.99\columnwidth]{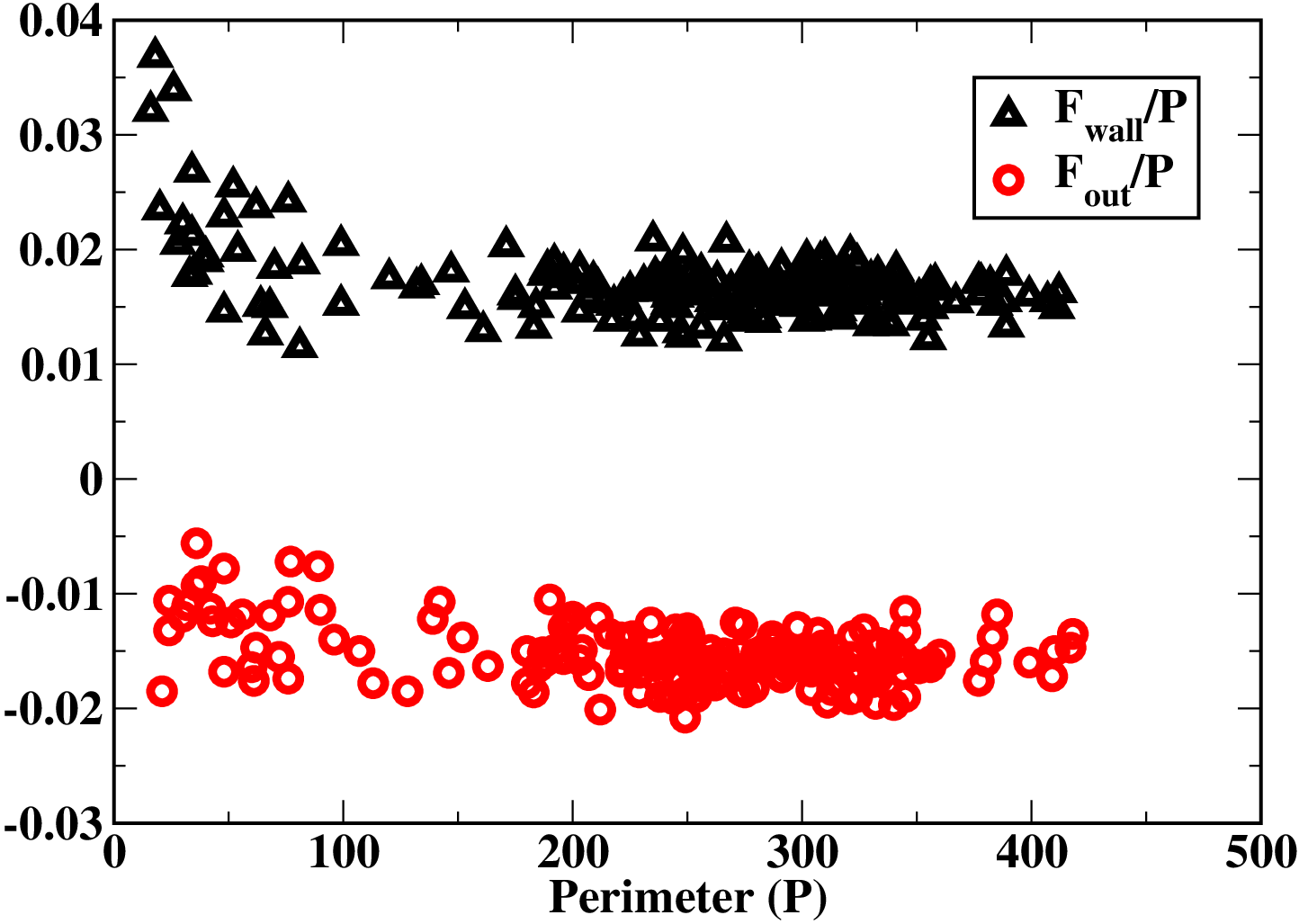}
\caption{(Color online) Random field fluctuation of the domain wall ($F_{wall}$) and just outside it ($F_{out}$) for $L=64$ at its critical disorder $W_{c}=0.265$. The y coordinate is the ratio $F/P$, and the x coordinate is P.} \label{FPwork}
\end{figure}

We have also calculated the ratio $F_{wall}/F$, where $F_{wall}$ is the random field energy on the domain wall. In Fig.~\ref{ratiowork}, one can see that the ratio is greater than $40\%$ for most of the configuration, which indicates that most of the random field energy is contained at the domain wall. We have also calculated ratio $F_{wall}/P$ and  $F_{out}/P$ where $F_{out}$ is the random-field energy of the sites just outside the domain wall. Our results as shown in Fig.~\ref{FPwork} suggests that $F$ is proportional to the perimeter of the domain and not its square root as assumed in Imry-Ma argument. So our results show that the various assumptions of the Imry and Ma picture are not valid in CG case. Similar conclusions were also drawn in the RFIM case \cite{Cambier,Esser}.
The calculated values of the exponents for different system sizes are summarized in Table~\ref{T1}.
\begin{table}[h]
\caption{\label{T1} Structural exponents for the 2D Coulomb glass.}
\begin{ruledtabular}
\begin{tabular}{ | l | l | l | l | }
     \hline
$L$ & $W_{c}$ & $\tau$ & $\lambda$  \\ \hline
16&0.35&0.6510&0.5237\\ \hline
32&0.30&0.6549&0.5916\\ \hline
48&0.275&0.6559&0.6292\\ \hline
64&0.265&0.7030&0.6896\\ \hline
\end{tabular}
\end{ruledtabular}
\end{table}
\section{Conclusion}
\label{sec6}
Two dimensional lattice model of CG at zero temperature was studied, where the minimum energy state was obtained by annealing the system using Monte Carlo simulation followed by cluster flipping. We have presented a finite size scaling analysis of the numerical data for systems upto $L=96$. Results suggests that a 2D CG with box type distribution of fields at $T=0$ exhibits a first-order transition at $W_{c}=0.2253$. The transition was characterized by the exponents $\nu=1.0$, $\beta=0$ and $\overline{\eta}=2.0$. The distribution of staggered magnetization at $W_{c}$ possessed three peaks which got sharper as system size was increased. The roughening argument given by Binder for RFIM is not satisfied for our system. The domain picture at $W_{c}$ is appreciably different from the one assumed by Imry and Ma. We found non-compact domains which were pinned at a certain location. Pinning was found to be independent of the initial spin configurations. Our calculations shows that most of the random field energy of the domain is contained in the domain wall. A two dimensional nearest neighbour Ising model on a triangular lattice in absence of disorder with all bonds antiferromagnetic has a large ground state degeneracy \cite{Wannier} due to geometrical frustrations. This large ground state degeneracy is immediately lifted for electrons on a quarter filled triangular lattice due to long range Coulomb interactions. The lifting of degeneracy is also accompanied by emergence of very many low lying metastable states with amorphous "stripe-glass" spatial structure \cite{Vladimir}. Extended Dynamical Mean Field calculations show that the ground state has stripe order and a first order transition is observed from liquid to stripe ordered phase as temperature is lowered \cite{Vla}. In our case, there is two fold degeneracy in COP in absence of disorder which is lifted by random fields. The COP ground state is now unique and competition  between interaction and disorder leads to phase transition at $W_{c}$ at $T=0$. Our method can be used to study disordered systems with long range interactions. One can also extend this work to CG with positional disorder. 
\section{Acknowledgement}
\label{sec7}
We thank late Professor Deepak Kumar for useful discussions on the subject. We wish to thank NMEICT cloud service provided by BAADAL team, cloud computing platform, IIT Delhi for the computational facility. Preeti Bhandari acknowledges UGC, Govt. of India for financial support through UGC-BSR fellowship (F.25-1/2013-14(BSR)/7-93/2007(BSR)).

\end{document}